\documentclass[aps,pra,twocolumn,superscriptaddress]{revtex4}
\usepackage{amsfonts}
\usepackage{amssymb}
\usepackage{amsmath}
\usepackage{epsfig}
\usepackage{color}
\usepackage{graphics, graphicx}
\usepackage{bbold}
\usepackage{psfrag}
\usepackage{mathcomp}
\usepackage{subfigure}
\usepackage{verbatim}
\usepackage{CJK}
\usepackage[colorlinks,citecolor=blue]{hyperref}

\begin{document}

%\preprint{}

\title{Entanglement Phase Transition in Chaotic non-Hermitian Systems}


\author{Zhen-Tao Zhang}
\email[]{zhzhentao@163.com}
\affiliation{School of Physics Science and Information Technology, Liaocheng University, Liaocheng, 252059, China}
\affiliation{Jiangsu Key Laboratory of Quantum Information Science and Technology, Nanjing University, Nanjing, 215163, China}
\author{Feng Mei}
\affiliation{State Key Laboratory of Quantum Optics Technologies and Devices, Institute of Laser Spectroscopy, Shanxi University, Taiyuan, Shanxi 030006, China}
\affiliation{Collaborative Innovation Center of Extreme Optics, Shanxi University, Taiyuan, Shanxi 030006, China}
%\affiliation{Frontier Science Center for Quantum Information, Beijing 100184, China}


\date{\today}

\begin{abstract}
  We study an entanglement phase transition in a class of chaotic non-Hermitian spin chains whose spin–spin coupling terms commute with the non-Hermitian contributions. Two representative models are investigated: the transverse-field Ising model with a complex longitudinal field and the non-Hermitian XX model with a transverse field. By analyzing their complex spectra, we find that both models undergo a dissipation-induced gapless–gapped phase transition when the transverse field exceeds a model-dependent threshold. Interestingly, the complex gap does not vary monotonically with the dissipation rate; instead, it exhibits pronounced oscillations before entering the gapped phase. By simulating their non-unitary dynamics, we show that the steady-state entanglement entropy undergoes a transition from volume-law to area-law scaling as the dissipation rate increases. Moreover, several unexpected features emerge within the volume-law regime: a larger complex gap or dissipation rate may lead to a more entangled steady state. We trace these unusual behaviors of the complex gap and the steady-state entanglement to level crossings between the maximal imaginary level and other spectral levels. Our work uncovers an exotic entanglement transition in chaotic non-Hermitian many-body systems.
 
\end{abstract}

%\keywords{}

\maketitle

\section{Introduction}
Entanglement in many-body quantum system plays a fundamental role in many fields of physics. For an open quantum system coupled to its surrounding environment, the entanglement between its subsystems may be destroyed, created or stabilized under the influence of dissipations. As we vary relevant parameters, the system would undergo phase transitions between distinct entanglement phases. Recently, one kind of such phenomenon referred to as measurement induced entanglement phase transition (MIPT) \cite{Li18,Li19,Skinner19,Choi20,Gullans20a,Gullans20b,Jian20,Bao20}, attracts wide attention in the communities of condensed matter physics and quantum information science. One typical platform to demonstrate MIPT is hybrid quantum circuit, in which nontrivial two-qubit unitary gates and single-qubit measurements are randomly applied to an array of qubit. At low measurement rate, the saturated entanglement is linearly scaled with system size, i.e., volume-law phase; once the measurement rate is larger than a critical value, the system enters area-law phase with the entanglement entropy independent of system size. The phenomenon can be understand by noticing that two-qubit unitary gates are able to generate entanglement while single-qubit measurements always suppress entanglement, thus the competition between them leads to a phase transition.\\
\indent Another setup for studying MIPT is monitored quantum many-body system. These include fermion systems \cite{Cao19,Buchhold21,Alberton21,Zhang22,Muller22,Yang23,Szyniszewski23,Gal23,Poboiko23,Kells23,Fava23,Merritt23,Poboiko24,Zhou24a,Soares25,Starchl25}, bosonic systems \cite{Delmonte25,Barberena25,Li25}, Majorana systems \cite{Sriram23,Lavasani23,Klocke23,Zhou24b} and so on. In this paper, we focus on one-dimensional spin chain systems. For an open quantum system coupled to a Markov environment, the state evolution follows the well-konwn Lindblad master equation. However, the density matrix obtained in this way is the averaged state over all measurement trajectories, and thus is blind to MIPT. To tackle this problem, we should evaluate the entanglement in single quantum trajectories. One regular trajectory is that of no-quantum jump, which can be realized via post-selection. In this case, the state evolution is governed by the Schrodinger equation with a non-Hermitian Hamiltonian. For a generic Ising chain with $\sigma^x$ measured and no-quantum jump selection, the Hamiltonian reads
\begin{eqnarray}\label{eq1}
	H=-J\sum\sigma^z_j\sigma^z_{j+1}-\Omega\sum_{j=1}^{N}\sigma_j^x-\frac{i\gamma}{4}\sum_{j=1}^{N}\sigma_j^x .
\end{eqnarray} 
In this integrable non-Hermitian model, if the coupling strength $J$ is larger than the transverse field $\Omega$, the system entanglement obeys a logarithmic law in size for low dissipation rates and area law for high dissipation rates. Therein, an entanglement phase transition occurs at the critical dissipation rate $\gamma_c=4\sqrt{J^2-\Omega^2}$. Take this typical model as a reference, one may wonder whether other kinds of entanglement transitions exist in integrable or chaotic non-Hermitian spin systems. Recently, many works are devoted to explore the entanglement phase transition in various non-Hermitian spin chain systems \cite{Gopalakrishnan21,Jian21,Biella21,Liu25,Malakar24,Muldoon23,Soares24,Turkeshi21,Turkeshi23,Xing24,Zhou25,Zhang25}. \\
\indent Among them, Ref. \cite{Jian21} has investigated the transverse field Ising model with a longitudinal complex field. They found a new kind of entanglement phase transition, which is triggered by Yang-Lee singularity. Along this line, we concentrate on a broad class of models, where the non-Hermitian term of the Hamiltonian commutes with the coupling term. To do this, we have studied the non-Hermitian XX model (NHXX) with a transverse field as well as the non-Hermitian transverse field Ising model (NHTFI). In these models, if the transverse field is vanishing, the imaginary part of their spectra is always gapped, consequently no entanglement transition. In the presence of the transverse field, these system are chaotic. Through numerical calculations of the complex spectra and entanglement entropy of the steady state, we find several unusual features of them. Firstly, an entanglement transition occurs in both models as the transverse field exceeds certain critical value, which is different for the considered models. Secondly, the complex gap in the volume-law phase is not a monotonous function of the dissipation rate. Instead, the gap could vary in oscillating patterns with the latter. Thirdly, the energy level with the maximal imaginary part may cross with other levels, which lead to abrupt jumps of the real part of its eigenenergy. As a result, the entanglement entropy of the steady state would jump downward or upward at the crossing points. Our findings provide an unconventional landscape of entanglement transition in chaotic non-Hermitian systems.\\
\indent The article is organized as follows. In Sec. \ref{sec2}, we introduce the numerical approach used to simulate the non-unitary dynamics of non-Hermitian systems. Sec. \ref{sec3} is devoted to studying the complex spectrum and the entanglement entropy of non-Hermitian transverse field Ising model. In Sec. \ref{sec4}, we investigate the complex gap and entanglement phase transition in the non-Hermitian XX model. After that, a detail discussion about the common features and differences between the two models is given in Sec. \ref{sec5}. We conclude this work in the last section.   
\section{Numerical simulation of non-unitary dynamics}\label{sec2}
  In this work, we consider open spin chain systems coupled with Markovian environments. The system could be subjected to decoherence channels or measurements, therefore, its state evolution is not unitary anymore. Even though, if we continuously monitor the decoherence or record the measurement results, the state of the system at each time is a pure state, which is conditioned on the collected results. One realization of such stochastic state evolution is called a quantum trajectory. For quantum trajectories that include quantum jumps, the state vector of the system evolves according to a stochastic Schrodinger equation of the form
  \begin{eqnarray}\label{eq2}
  	d|\psi(t)\rangle=-idt[H-\frac{i}{2}\sum_{j}(\hat{L}_j^\dagger\hat{L}_j-\langle\hat{L}_j^\dagger\hat{L}_j\rangle_t)]|\psi(t)\rangle\nonumber\\+\sum_{j}(\frac{\hat{L}_j}{\sqrt{\langle\hat{L}_j^\dagger\hat{L}_j\rangle_t}}-1)dN_j(t)|\psi(t)\rangle,
  \end{eqnarray}
  where $\hat{L}_j$ denotes jump operator of the spin at the site j, $\langle\hat{L}_j^\dagger\hat{L}_j\rangle_t$ is the expected value of the operator $\hat{L}_j^\dagger\hat{L}_j$ in the conditioned state $|\psi(t)\rangle$. $dN_j(t)\in\{0,1\} $ are independent random variables following Possion distribution with the average value  
  $\overline{dN_j(t)}=\langle\hat{L}_j^\dagger\hat{L}_j\rangle_tdt$.\\
  \indent  The state evolution in a single step is either continuous non-unitary evolution driven by a non-Hermitian Hamiltonian, or a disrupt quantum jump given by $\hat{L}_j|\psi(t)\rangle$. If one post-selects the quantum trajectory without any quantum jump, its dynamics is determined by the non-Hermitian Hamiltonian
  \begin{eqnarray}\label{eq3}
  	H_{NH}=H-\frac{i}{2}\sum_{j=1}\hat{L}_j^\dagger\hat{L}_j.
  \end{eqnarray}
  Instead, if all the trajectories are taken into account, the unconditioned density matrix $\rho$ of the system is obtained by averaging the conditional density matrix $\rho_c=|\psi(t)\rangle\langle\psi(t)|$ over all measure records. In this case, the density matrix evolves according to the Lindblad master equation:
   \begin{eqnarray}\label{eq4}
   \frac{d\hat{\rho}(t)}{dt}=-i[\hat{H},\hat{\rho}]+\sum\limits_{j}(\hat{L}_j \hat{\rho} \hat{L}_j^{\dagger}-\frac{1}{2}\{\hat{L}_j^{\dagger}\hat{L}_j, \hat{\rho}\}).
   \end{eqnarray}                                              
Here, we focus on the dynamics of the system following the specific trajectory without quantum jump. In this situation, the state evolution is governed by the Schrodinger equation with a non-Hermitian Hamiltonian (\ref{eq3}).\\
\indent To simulate the non-unitary dynamics, we adopt the newly introduced Faber polynomial method \cite{Soares24}. For a time-independent Hamiltonian, the time evolution operator of the state is given by
   \begin{eqnarray}\label{eq5}
 U=e^{-iH_{NH}t}.
 \end{eqnarray}   
It has been shown that the operator $U$ can be expanded with a Faber polynomial,
 \begin{eqnarray}\label{eq6}
 	U=\sum_{n=0}^{\infty}c_n(t)F_n(\tilde{H}),
 \end{eqnarray} 
where $F_n$is the nth Faber polynomial, and $\tilde{H}=H_{NH}/\lambda$ is the rescaled Hamiltonian ($\lambda$ is obtained from the bounds of the real and imaginary part of the spectra of the Hamiltonian $H_{NH}$). Using this method, we could efficiently simulate the state evolution with the precision controlled by the truncated order of the Faber polynomial. In following calculations, we truncate the polynomial up to the order of 5, which is great enough for the time step we choose. Since the operator $U$ is not unitary, the state vector after the evolution in
 each step $\delta t$ should be normalized such that
 \begin{eqnarray}\label{eq7}
	|\psi(t+\delta t)\rangle=\frac{U|\psi(t)\rangle}{||U|\psi(t)\rangle||}.
\end{eqnarray} 
Consequently, we can obtain the normalized state of the quantum many-body system at any time. \\
\indent In this paper, we use the entanglement entropy to measure the bipartite entanglement in a spin chain with $N$ sites. For this sake, we partition the system into two halves, part A and part B (with equal chain length, $N_A=N_B=N/2$). The reduced density matrix of part A can be obtained by partial tracing out the degrees of freedom of part B: $\rho_A=\text{Tr}_B(|\psi\rangle\langle\psi|)$. Then, the entanglement entropy is given by
\begin{eqnarray} \label{eq8}
	S=-\text{Tr}_A(\rho_A\text{ln}\rho_A)
\end{eqnarray}
In latter sections, we employ this formula to calculate the entanglement entropy.
\section{Entanglement In non-Hermitian transverse field Ising model }\label{sec3}
The first model we consider is the transverse field Ising model with longitudinal non-Hermitian terms. Its Hamiltonian can be written as:
\begin{eqnarray}\label{eq9}
	H_{NHTFI}=-J\sum\sigma^z_j\sigma^z_{j+1}-\Omega\sum_{j=1}^{N}\sigma_j^x-\frac{i\gamma}{4}\sum_{j=1}^{N}\sigma_j^z ,
\end{eqnarray} 
where $\sigma_j^x,\sigma_j^y,\sigma_j^z$ are Pauli operators of the spin at site $j$, and $J$ denotes nearest-neighbor coupling strength. $\gamma$ is the homogeneous dissipation rate of each spin. Notice that the non-Hermitian term $-\frac{i\gamma}{4}\sum_{j=1}^{N}\sigma_j^z$ is different from that in Eq. (\ref{eq1}). Henceforth, we focus on the case with periodic boundary condition, and set the coupling strength $J=1$. \\
\subsection{phase transition between gapless phase and gaped phase}
In the Hamiltonian (\ref{eq9}), the non-Hermitian term commutes with the spin-spin coupling term. Thus, if $\Omega=0$, the imaginary part of the energy spectra is always gaped with the gap $\Delta=\gamma/2$, where $\Delta$ denotes difference between the
largest and second-largest imaginary part of the spectra. The presence of a transverse field makes the model chaotic. Ref. \cite{Malakar24} has shown that the model is still gaped as long as $\Omega<J$. Consequently, there is no entanglement phase transition in this regime. Here, we mainly investigate th spectrum of the system in the opposite regime, i.e., $\Omega\geq J$. \\
\begin{figure}
	\includegraphics[width=8.8cm]{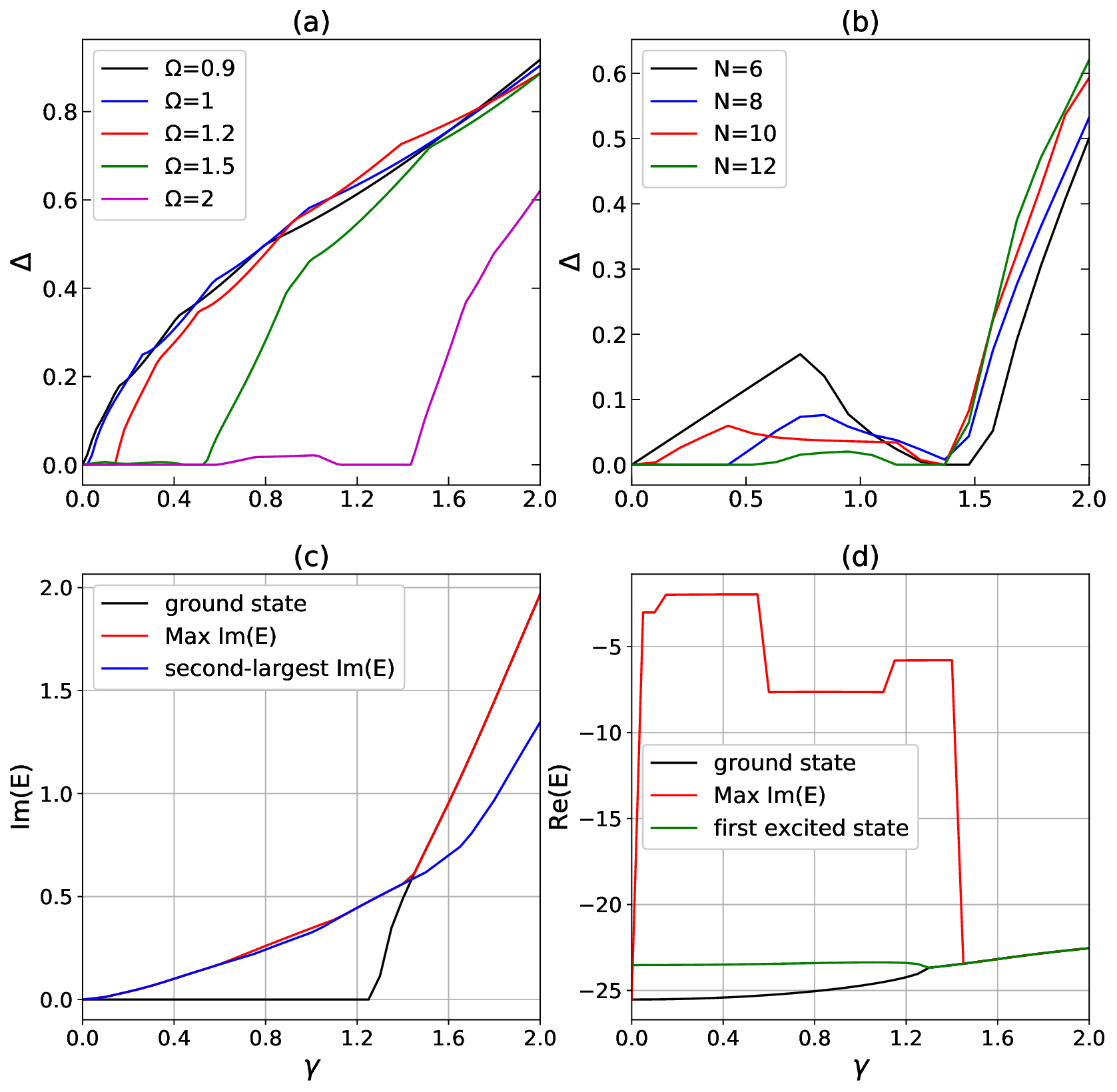}
	\caption{Complex gap and spectra of NHTFI model with respect to $\gamma$. (a) The gap for different $\Omega$ with the chain length N=12. (b) The gap for different N with $\Omega=2$. The imaginary part (c) and real part (d) of eigenenergies of several levels with N=12 and $\Omega=2$. \label{fig1}}
\end{figure}
\indent We have numerically calculated the complex gap and spectra of eigenenergies of several levels (Fig. 1). The gap of a spin chain of N=12 with the parameter $\Omega$ on the range $[0.9, 2]$ is plotted in Fig. 1(a). As the transverse field strength is 0.9, the gap is consistently lifted for any non-vanishing $\gamma$. The situation starts to change once the field increases up to 1. We can see that in the regime $\Omega>1$, there merges two distinct phase: gapless phase for small $\gamma$ and gapped phase for large $\gamma$. Between these two phase locates a phase transition point with the critical dissipation rate $\gamma_c$ smaller than $4\Omega$. Moreover, the quantity $\gamma_c$ increases with $\Omega$, and gradually converges to $4\Omega$ in the large field limit (not shown in the figure). In fact, if $\Omega$ is much larger than the coupling strength, the system can be considered approximately as separate non-Hermitian spins, whose critical point of the dissipation rate between gapless-gapped phase is located at $4\Omega$. In a word, the NHTFI model features a sharp gapless-gapped phase transition as the transverse field is larger than the coupling strength. \\
\indent It is worth noting that the gap in the gapless phase is not necessarily vanishing for finite chain length. In Fig. 1(a), we can see for $\Omega>1$ the gap manifests weak oscillations in the low dissipation rate region. In detail, the variation of the gap with $\gamma$ is neither monotonous nor smooth, which differs it from non-Hermitian spin chain models studied before \cite{Gopalakrishnan21,Malakar24}. At $\Omega=2$, we have also calculated the gap of spin chains with various length N, as shown in Fig. 1(b). We can see that the gap behave similarly and share a common critical dissipation rate. Moreover, in the gapless phase, the gap oscillates with an amplitude that is suppressed with the increasing of N. \\
\indent To understand the unusual  properties of the system, we have calculated the imaginary and real part of eigenenergies of several typical levels for N=12 and $\Omega=2$, shown in Fig. 1(c)-(d). Note that the ground state is the level with lowest real part of the eigenenergy. Before entering the gapped phase, the real part of the maximal imaginary level is far above that of the ground state. These levels cross together at the critical point $\gamma_c=1.45$, after which the ground state take over the maximal imaginary level (Fig. 1(c)). Therefore, the real part of the latter jumps down at the critical dissipation rate (Fig. 1(d)). The gapless and gapped phase transition is triggered by Yang-Lee singularity, where the imaginary part of the ground state rises up from zero \cite{Jian21}. Actually, within the gapless phase there are more crossings between the maximal imaginary level and other levels, marked by jumps of real part of the former. These level crossings lead to the oscillating gap before the phase transition. 
\subsection{entanglement dynamic and entanglement phase transition}
\begin{figure}
	\includegraphics[width=8.5cm]{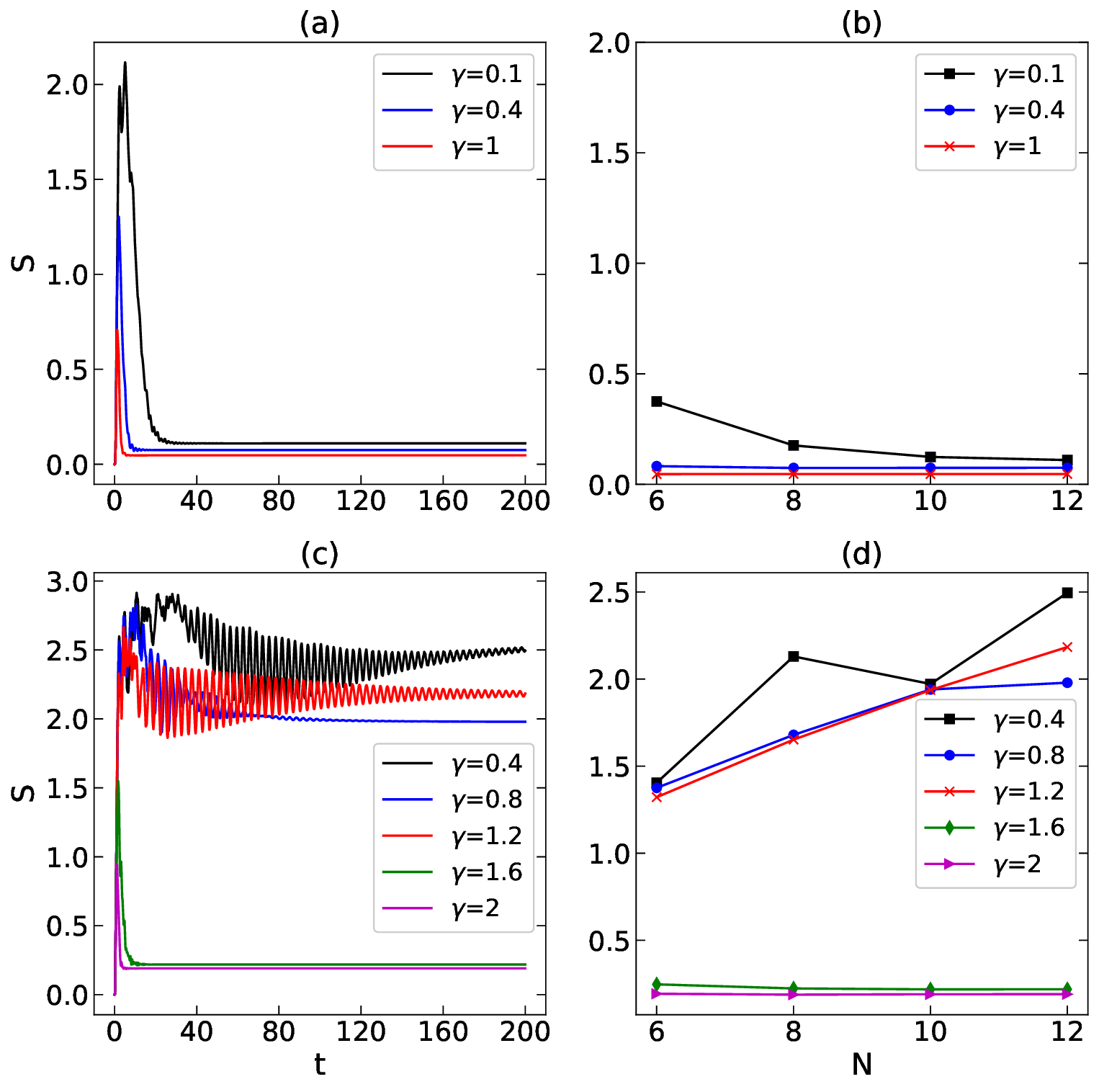}
	\caption{ Entanglement dynamics of the NHTFI model and entanglement entropy of the steady state in N, with $\Omega=0.9$ in (a)-(b), and $\Omega=2$ in (c)-(d). \label{fig2}}
\end{figure}

Entanglement dynamics of the chain is tightly related with the complex gap of the system, and the long-time steady state would be the maximal imaginary level. To check the existence of entanglement transition, we have calculated the half-cut entanglement entropy $S$ using the method introduced in the former section.\\
\indent In Fig. \ref{fig2}, we present the evolution of the entanglement entropy for a chain with length $N=12$ and its steady value for different chain lengths. Without loss generality, the initial state of the system is set to be the separate state $[(|0\rangle+|1\rangle)/\sqrt{2}]^{\otimes N}$. As $\Omega=0.9$, the entanglement could quickly evolve to a stable low level (Fig. 2(a)), even for the dissipation rate $\gamma$ much smaller than $\Omega$. Moreover, the final entanglement entropy is basically independent of the chain length for any finite dissipation (Fig. 2(b)). These results prove that in the regime $\Omega<1$ there is no entanglement phase transition, which agrees with previous studies \cite{Malakar24}. Alternatively, when $\Omega=2$, the entanglement entropy evolves in two different manners (Fig. 2(c)). As $\gamma<\gamma_c=1.45$ the entropy oscillates at a high level for a long time before stabilization. On the contrary, as $\gamma>\gamma_c=1.45$ the entanglement entropy would quickly decay to a small value. Accordingly, the scaling of the entanglement in size is distinct in these two regions, as shown in Fig. 2(d). In the gapless phase, the entanglement entropy is an extensive quantity, basically following a volume law; in the gapped phase, the entanglement is independent of the chain length, indicating an area law scaling.\\
\indent Generally, it is expected that the complex gap and the steady-state entanglement of non-Hermitian systems are both monotonous functions of local dissipation rates, i.e., stronger dissipation would give rise to larger gap and less entangled steady state \cite{Gopalakrishnan21,Malakar24}. However, we find exceptional cases in the model considered. As shown in Fig. 2(c), the time taken to reach the saturated entanglement entropy is longer for $\gamma=1.2$ than that for $\gamma=0.8$. This anomalous result is due to the relative larger gap at $\gamma=0.8$ (see Fig. 1(a)). More interestingly, the steady value of entanglement entropy is also reversed at these two points, as illustrated in Fig. 2(c)-(d). We attribute this anomaly to the level crossing located at $\gamma=1.1$ (see Fig. 1(c-d)).  Since the level crossing is accompanied with a switch of the maximal imaginary level, the entanglement entropy of the steady state would jump simultaneously. Therefore, the level crossings in the volume-law phase would lead to unconventional features of the system.\
\indent 
\section{Entanglement In non-Hermitian xx model} \label{sec4}
\subsection{in the absence of transverse field}
We start by analyzing the NHXX model with vanishing transverse field. The Hamiltonian reads:
\begin{eqnarray}\label{eq10}
	H_{XX}=-J\sum(\sigma^x_j\sigma^x_{j+1}+\sigma^y_j\sigma^y_{j+1})-\frac{i\gamma}{4}\sum_{j=1}^{N}\sigma_j^z ,
\end{eqnarray}
where $J$ denotes nearest-neighbor transverse coupling strength. We can see that the first term of the Hamiltonian $H_{XX}$ is number-conserving. Moreover, if the state of the spin chain is initialized in a subspace of fixed excitation number, the non-Hermitian term $-\frac{i\gamma}{4}\sum_{j=1}^{N}\sigma_z$ has little effect on the system dynamics since it is just an imaginary constant within the subspace. In this case, the physical meaning of the non-Hermitian term is that the system is continuously monitored and only the quantum trajectory of no-quantum jumps is post-selected. Except these operations, the dynamic of system localized in a number-conserving subspace is equal to the unitary evolution governed by the Hermitian part of the Hamiltonian. Alternatively, if the initial state is a superposition of eigenstates of the total excitation operator $\sum_{j=1}^{N}\sigma_z$, the evolution of the system is not unitary. The reason is that the non-Hermitian term appears in the Hamiltonian as imaginary diagonal elements, which is divergent for different excitation subspaces.\\
\begin{figure} 
	\includegraphics[width=8cm]{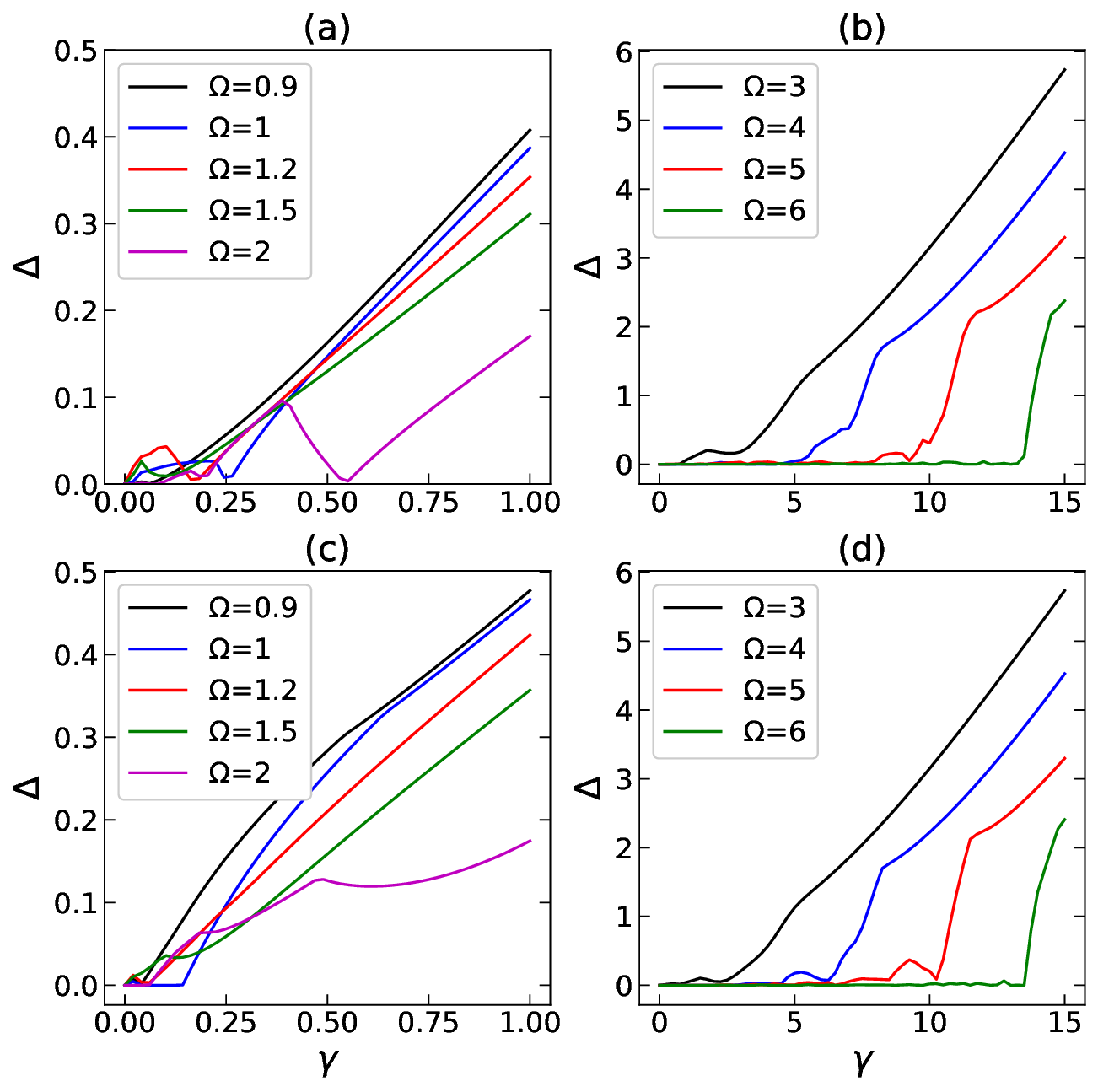}
	\caption{Complex spectral gap with respect to $\gamma$ in NHXX model. The chain length N=10 in (a)-(b), and N=12 in (c)-(d). The values of $\Omega$ are listed in the legend of each figure.  \label{fig3}}
\end{figure}
\indent Now we consider the entanglement produced by the Hamiltonian $H_{XX}$. First we assume that the system is prepared at a separate state with fixed total excitation number. In this situation, the entanglement entropy would oscillate periodically between zero and its maximum. Furthermore, the maximal entanglement entropy relies on the excitation number. For a chain with N spins, the maximal entanglement is distributed symmetrically centered at excitation number $n=N/2$ due to the $Z_2$ spin-inversion  symmetry of the Hermitian part of the Hamiltonian. The scaling of the entanglement entropy with the system size $N$ is determined by $n$. For example, if $n=1$ for any chain length, the entanglement is constant for different $N$. Instead, if the initial state is a N\'{e}el state $|0,1,...,0,1\rangle$, i.e., $n=N/2$, the maximum of the entanglement scales linearly with the size $N$. For a generic initial state,  the population of the state within the excitation $n$ subspace decays exponentially as
\begin{eqnarray}\label{eq11}
	p_n(t) = p_n(0)e^{-n\gamma t} .
\end{eqnarray}
Therefore, after a long enough time only the basis states with least excitation survives, and finally the system follows an unitary evolution. At this stage, the entanglement entropy is determined by the least excitation number of the initial state. Therefore, the steady-state entanglement totally relies on the initial state.
\subsection{In the presence of a transverse field}
We turn to investigate the entanglement phase transition in the presence of a transverse field. In this case, the total Hamiltonian of the system is given by 
\begin{eqnarray}\label{eq12}
	H_t=H_{XX}-\Omega\sum_{j=1}^{N}\sigma_j^x,
\end{eqnarray}
where $\Omega$ is the strength of the homogeneous transverse field. Since the transversal field term does not commute with either the coupling term or the non-Hermitian term in the Hamiltonian, it thoroughly breaks the excitation number conservation of the spin chain. Thus, the state of the system would not be constrained in any excitation subspace anymore. From the view of entanglement generation, the coupling term is an indispensable part, and the non-Hermitian term would minimize the excitation, which is not beneficent for producing entanglement. In this context, the effect of the transverse field on the entanglement is elusive. As all the three terms in the Hamiltonian (\ref{eq12}) are finite, the system is chaotic. Thus, it is difficult to get analytical results about its spectrum and entanglement phase transition. Therefore, we make use of the numerical approach introduced in Sec. \ref{sec2} to address this problem.\\

\subsubsection{complex gap}
\indent We have calculated the gap of the imaginary part of the eigenvalues of the total Hamiltonian $H_t$ with $\gamma$ for N=10 (Fig. 3(a)-(b)), and for N=12 (Fig. 3(c)-(d)). Fig. 3(a) and (c) are taken for modest transverse field $\Omega\in[0.9, 2]$, while Fig. 3(b) and (d) are illustrating the gap for $\Omega$ on the range $[3, 6]$. We can see that as $\Omega=0.9$, a gapless phase appears in low $\gamma$ region for both chain sizes. This indicates that the gapless-gapped phase transition starts to merge as the transverse field is smaller than the coupling strength, which is distinct from NHTFI model studied in the former section. Another remarkable feature of the gap in modest $\Omega$ regime is that the critical point is not very clear. As shown in Fig. 3(a),(c), the variation of the gap is irregular, without an unique pattern for different $\Omega$ and N. Therefore, it is hard to pinpoint a gapless-gapped phase transition. However, with the increasing of $\Omega$, the crossover between gapless and gapped phase gradually converges to a sharp transition at the common critical dissipation rate $\gamma_c=4\Omega$ for either chain length (see Fig. 3(b),(d)).\\
\begin{figure}
	\includegraphics[width=8.8cm]{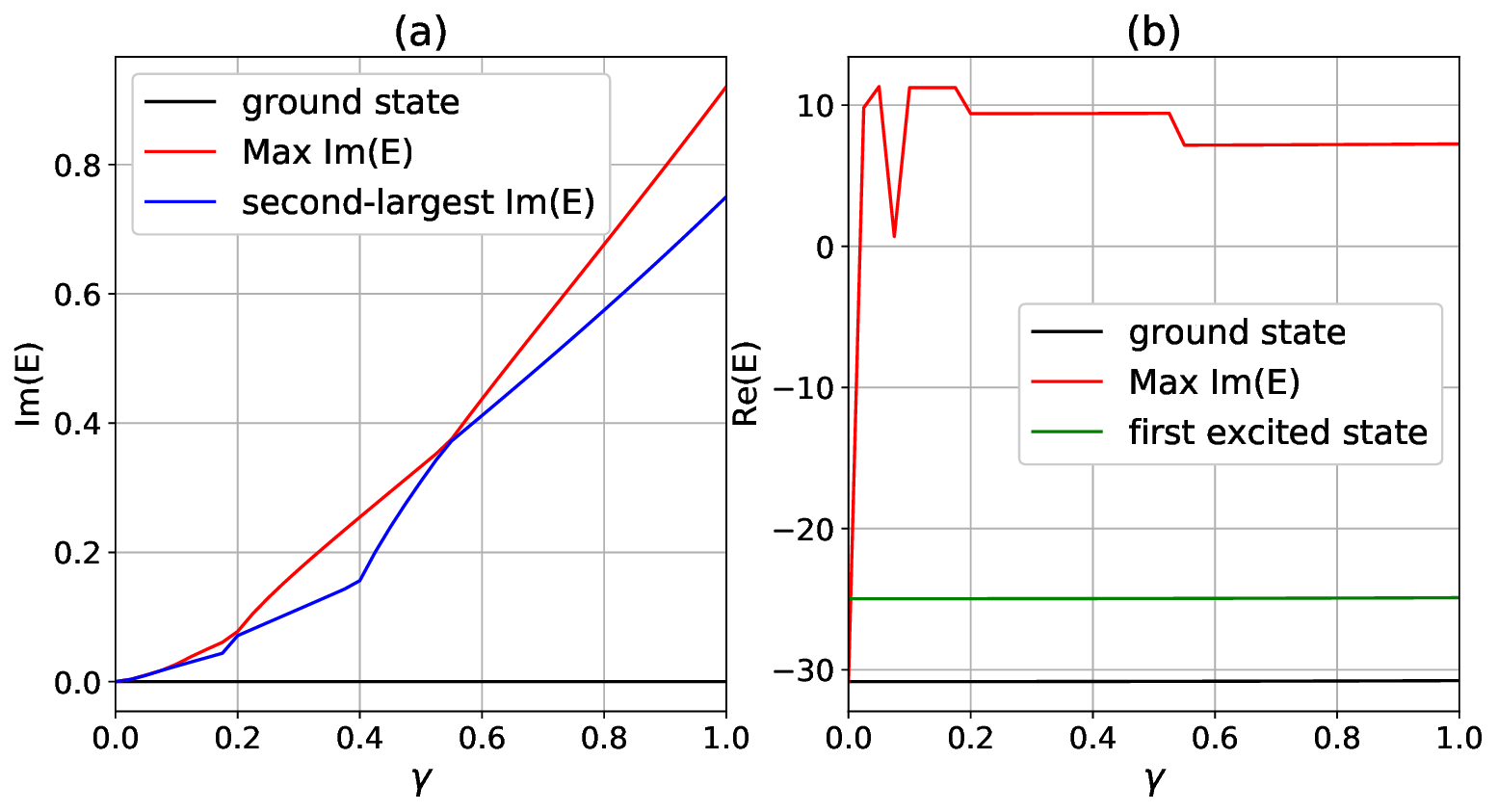}
	\caption{ The imaginary part (a) and real part (b) of the eigenenergies with respect to $\gamma$ for chain length N=10. The value of the parameter $\Omega$ is set to be $2$. \label{fig4}}
\end{figure}
\indent Now we analyze the variation of the gap in the low dissipation region and its underlying mechanism. As shown in Fig. 3, the gap varies with the dissipation rate similarly with NHTFI model: it is not a smooth function of the dissipation rate, and manifests some oscillations. To understand these phenomena, we have calculated the imaginary and real part of eigenenergies of relevant levels for N=10 and $\Omega=2$, as shown in Fig. 4. We can see that the maximal imaginary level crosses with other levels more than one time, which lead the real part of the former switches suddenly at these crossing points. This also appears in NHTFI model. However, the maximal imaginary level in the gapped phase is distinct for the two models. Remember that the ground state of NHTFI model takes over the maximal imaginary level after phase transition. In contrast, here the ground state retain zero imaginary part, and the real part of the maximal imaginary level is much larger than the ground state. This result implies that there is no Yang-Lee singularity in the current model. Actually, we can see from Fig. 4(b) that the ground state is not degenerate on whole range of the dissipation rate, which is different from the case of NHTFI model (shown in Fig. 1(d)). 
\subsubsection{entanglement phase transition}
Before studying the entanglement scaling in size, we first compare the gap for various chain lengths. The gap with the dissipation rate for $N=\{8,10,12,14\}$ are illustrated in Fig. 5(a, c) with $\Omega=0.9$, 2, respectively. We can see that the critical point between gapless and gapped phase varies with the system size, which is obviously distinct with NHTFI model. Moreover, for a fixed dissipation rate, the gap is not monotonously suppressed with the increasing of system size. For instance, as $\Omega=0.9$, the gap of the chain with $N=12$ exceeds that of the chain with N=10 on the whole range of $\gamma$.  \\
\begin{figure}
	\includegraphics[width=8.5cm]{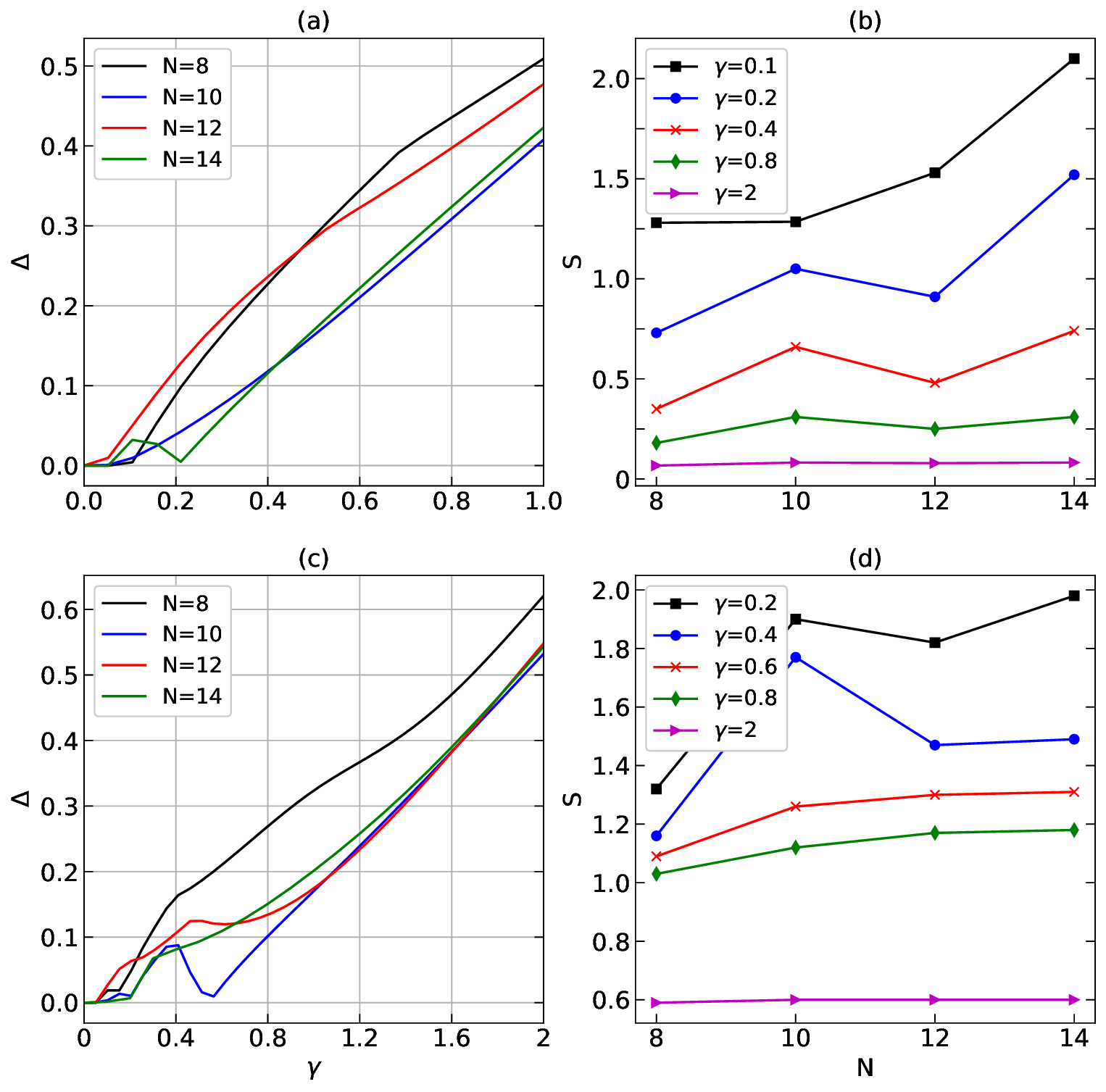}
	\caption{Complex gap with respect to $\gamma$ for different chain length and the entanglement entropy of the steady state as a function of N. In (a)-(b) the parameter $\Omega=0.9$ while $\Omega=2$ in (c)-(d). \label{fig5}}
\end{figure}
\indent To check the existence of entanglement phase transitions, we have calculated the steady value of entanglement entropy with chain length for various $\gamma$, as shown in Fig. 5(b, d). When $\Omega=0.9$ (Fig. 5(b)), the entanglement entropy ascends linearly with N for small $\gamma$ except some fluctuations due to finite effect, which indicates a volume law entanglement phase. With the increasing of $\gamma$, the growing of the entanglement entropy in size is gradually suppressed. As the dissipation rate is larger than 0.8, the entanglement is basically independent of the chain length, showing an area law phase. These results prove that the entanglement phase transition emerges in this model even if the transverse field is smaller than the coupling strength, which is not the case for NHTFI model (see Fig. 2(b)). As $\Omega=2$, the entanglement phase transition is maintained with a larger critical $\gamma$ (Fig. 5(d)). More interestingly, we find the entanglement entropy for $\gamma=0.4$ is much larger than that for $\gamma=0.6$, although the gap of the former exceeds the latter. This counter-intuitive phenomenon can be attributed to the level crossing between $\gamma=0.4$ and 0.6 (see Fig. 4). At the crossing point, the maximal imaginary level is taken over by another level with different real part of eigenenergy. As a result, the entanglement entropy of steady state jump downward suddenly. 
\section{Discussion}\label{sec5}
Our results have shown that an entanglement phase transition occurs under proper situation for both NHTFI model and NHXX model. Now we discuss the common features and differences of the two models as regards their complex spectra and steady-state entanglement.
Note that the spin-spin coupling term in Hamiltonian Eq.(\ref{eq9}) and Eq.(\ref{eq12}), either $-J\sum\sigma^z_j\sigma^z_{j+1}$ or $-J\sum(\sigma^x_j\sigma^x_{j+1}+\sigma^x_j\sigma^x_{j+1})$, commutes with the corresponding non-Hermitian term $-\frac{i\gamma}{4}\sum_{j=1}^{N}\sigma_j^z $. In the absence of transverse field, the imaginary part of their eigenenergies is always gapped for any finite dissipation rate $\gamma$, and the steady state (maximal imaginary level) of the systems is a product state. Therefore, neither gapless-gapped spectral transition nor entanglement phase transition exist in this case. Alternatively, if the coupling term is vanishing, these models degrade to separate non-Hermitian spins. In this case, due to parity-time symmetry of the uncoupled Hamiltonian, the overall complex spectra is subject to a gapless-gapped transition as the dissipation rate is tuned upward from zero. Of course, there is no entanglement between any pair of single spins. \\
\indent In general cases with parameters ${J, \Omega, \gamma}$ finite, the interplay of the coupling, the transverse field and the dissipation produces chaotic spectra of this kind of systems. When $\Omega$ is smaller than a threshold $\Omega_{Th}$, the imaginary part of their complex spectra are still gapped for any finite dissipation rate $\gamma$, and the steady-state entanglement follows the area law uniquely. The value of the threshold $\Omega_{Th}$ is model-dependent, which have been illustrated in the former sections. For NHTFI model, $\Omega_{Th}=J$, which is exactly the critical point between ferromagnetic phase and paramagnetic phase of the transverse field Ising model \cite{Jian21,Malakar24}. For NHXX model, it is hard to theoretically analyze the value of $\Omega_{Th}$. However, we can estimate it from the pattern of complex gap for different $\Omega$. As shown in Fig. 3(a),(c)), $\Omega_{Th}$ is a little lower than $0.9J$.\\
\indent When $\Omega>\Omega_{Th}$, these models permit two distinct phases: gapless phase and gapped phase. We emphasize that in the modest $\Omega$ regime ($\Omega \gtrsim J$), the complex gap in gapless phase is not always zero, instead it manifests some oscillations with the dissipation rate before entering the gapped phase. As illustrated in former sections, this phenomenon stems from level crossings between the maximal imaginary level and the remaining levels, which is a common feature of the models considered. Importantly, these spectral properties determine the entanglement structure of the steady state, leading to the volume-law scaling of the entanglement entropy in gapless phase. On other side, in gapped phase, the strong dissipation induces a low entangled steady state, following area-law scaling in size. Therefore, in both models the volume-law to area-law entanglement transition occurs in the regime $\Omega \gtrsim J$. Meanwhile, there are some remarkable differences between NHTFI model and NHXX model in the variation of their complex spectra, as illustrated in Fig. 1 and Fig. 3-4. The major distinction of the spectra is that the former harbors Yang-Lee singularity in the gapless phase and the maximal imaginary level in gapped phase is just the ground state. In NHXX model, there is no Yang-Lee singularity; as such, the maximal imaginary level does not coincide with its ground state. Before ending this section, we would like to briefly discuss the regime $\Omega\gg J$, i.e., the transverse field strength is much larger than the coupling strength. In this case, the complex gap is basically zero in the gapless phase, which converges to that of the separate non-Hermitian spins. Henceforth, the steady state is not well defined in this regime, and the entanglement entropy will oscillate constantly. Therefore, if we want to observe the steady-state entanglement phase transition in practice, the transverse field should not be much stronger than the spin-spin coupling.
\section{conclusion}
 In summary, we have studied two chaotic non-Hermitian spin chain models, in which the non-Hermitian term commutes with the coupling term. In both models, once the transverse field is larger than a model-dependent value, the systems would transition from gapless phase to gapped phase with the increase of the dissipation rate, which is accompanied with entanglement phase transition. In the gapless phase, we find some unusual features in the two models. First, the complex gap varies with the dissipation rate in an non-smooth oscillating manner; second, the maximal imaginary level is subject to crossings with other levels, which results in the real part of the former jumping upward or downward suddenly. Furthermore, these features of the complex spectra lead to unexpected results about their steady state: for certain chain length, larger dissipation rate or complex gap yields greater entanglement entropy of the steady state.\\
\indent Our work reveals an unconventional entanglement phase transition in chaotic non-Hermitian systems. Compared with other entanglement transitions, the most remarkable feature of it is the appearance of crossings between the maximal imaginary level and the remaining levels in the gapless phase. It generalizes the entanglement phase transition triggered by Yang-Lee singularity to a broader class of systems, in the sense that its maximal imaginary level in the gapped phase is not necessarily the ground state. In the future, we look forward to observe this kind of entanglement transition in a practical system.\\

\section*{ACKNOWLEDGMENTS}
This work is funded by the Introduction and Cultivation Plan of Youth Innovation Talents for Universities of Shandong Province, the National Key Research and Development Program of China (Grant No. 2022YFA1404201) and National Natural Science Foundation of China (Grant No. 12474361).

% If in two-column mode, this environment will change to single-column
% format so that long equations can be displayed. Use
%\begin{widetext}
%\end{widetext}

% Surround table environment with turnpage environment for landscape
% table
% \begin{table}
% \begin{ruledtabular}
% \end{tabular}
% \end{ruledtabular}
% \end{table}

%\appendix
%\section{}

%\begin{acknowledgments}
%\end{acknowledgments}


\end{document}